\documentclass[%
 aip,
 jmp,%
 amsmath,amssymb,
reprint,%
unsortedaddress
]{revtex4-1}

\usepackage{graphicx}
\usepackage{dcolumn}
\usepackage{bm}

\usepackage{epsfig}
\usepackage{color}
\usepackage{amsmath}
\usepackage{amssymb}
\usepackage{psfrag}
\usepackage{pstricks}
\usepackage{longtable}
\usepackage{dcolumn}
\usepackage{bm}

\providecommand{\abs}[1]{\lvert#1\rvert}

\begin{document}

\preprint{AIP/123-QED}

\title[Envelope molecular-orbital theory of extended systems I]{Envelope molecular-orbital theory of extended systems I.\\
Electronic states of organic quasi-linear nano-heterostructures}

\author{J.C. Arce}
\email{jularce@univalle.edu.co}
\affiliation{Departamento de Qu\'imica, Universidad del Valle, A.A. 25360, Cali, Colombia}
\author{A. Perdomo-Ortiz}
\altaffiliation[Present address: ]{Department of Chemistry and Chemical Biology, Harvard University, Cambridge, MA 02138.}
\affiliation{Departamento de Qu\'imica, Universidad del Valle, A.A. 25360, Cali, Colombia}
\author{M.L. Zambrano}
\affiliation{Centro de Investigaci\'on en Ciencias B\'asicas y Ambientales, Universidad Santiago de Cali, A.A. 4102, Cali, Colombia}
\author {C. Mujica-Mart\'inez}
\altaffiliation[Present address: ]{Institut f\"ur Theoretische Physik, Universit\"at Hamburg, Jungiusstrasse 9 20355, Hamburg, Germany.}
\affiliation{Departamento de Qu\'imica, Universidad del Valle, A.A. 25360, Cali, Colombia}

\date{\today}

\begin{abstract}
A conceptually appealing and computationally economical course-grained molecular-orbital (MO) theory for extended quasi-linear molecular
heterostructures is presented. The formalism, which is based on a straightforward adaptation, by including explicitly the vacuum, of the
envelope-function approximation widely employed in solid-state physics, leads to a mapping of the three-dimensional single-particle eigenvalue
equations into simple one-dimensional hole and electron Schr\"odinger-like equations with piecewise-constant effective potentials and masses. The
eigenfunctions of these equations are envelope MO's in which the short-wavelength oscillations present in the full MO's, associated with the
atomistic details of the molecular potential, are smoothed out automatically. The approach is illustrated by calculating the envelope MO's of
high-lying occupied and low-lying virtual $\pi$ states in prototypical nanometric heterostructures constituted by oligomers of
polyacetylene and polydiacetylene. Comparison with atomistic electronic-structure calculations reveals that the envelope-MO energies agree very
well with the energies of the $\pi$ MO's and that the envelope MO's describe precisely the long-wavelength variations of the $\pi$ MO's. This envelope
MO theory, which is generalizable to extended systems of any dimensionality, is seen to provide a useful tool
for the qualitative interpretation and quantitative prediction of the single-particle quantum states in mesoscopic molecular structures and the
design of nanometric molecular devices with tailored energy levels and wavefunctions.
\end{abstract}

\maketitle

\section{\label{sec:introduction} Introduction}

Extended molecular structures appear in condensed-matter physics, macro- and supra-molecular chemistry, molecular biology, materials science, molecular electronics and optoelectronics, and nanotechnology. Ranging from one-dimensional to three-dimensional, these include, among many others, inorganic, organic and biological polymers\cite{Ladik88,Shapir08} and wires;\cite{Dubin06} dendrimers;\cite{Lupton02} surfaces\cite{Hoffmann88} and graphene-based structures;\cite{Rao09} fullerenes\cite{Andreoni98} and nanotubes;\cite{Odom00} clusters;\cite{Catlow10} and crystals.\cite{Hoffmann88,Yu99} In addition, there exist assemblages of different fragments or materials, like hybrid structures\cite{Moorlag05} and heterostructures.\cite{Davies98}

When such an extended system is sufficiently large so that, for practical purposes, it can be considered as infinite, and periodic, so that a repeat unit cell can be identified, periodic boundary conditions can be employed, effectively reducing the calculation of its electronic structure to the one of the unit cell at a set of sampling \textit{k} points. Within (Schr\"odinger or Kohn-Sham) orbital approaches, this leads to a description in terms of energy bands and (Bloch) crystalline orbitals.\cite{Ladik88,Hoffmann88,Yu99} On the other hand, if the system lacks translational symmetry alternative approaches must be sought.

Non-periodic extended molecular structures of interest in the abovementioned fields can possess from hundreds up to millions of atoms. Depending on the degree of accuracy required and the computational resources available, there are semi-empirical\cite{DiCarlo03} and ab-initio\cite{Ladik99} electronic-structure methods to choose from. When coupled with linear-scaling algorithms,\cite{Goedecker99} these methods can handle large numbers of atoms,\cite{Yu09} like the impressive recent million-atom implementation of orbital-free DFT.\cite{Hung09} Other approaches, like density-matrix renormalization theory,\cite{Schollwock05} elongation methods,\cite{Imamura91} fragment molecular-orbital theory,\cite{Kitaura99} and embedded cluster methods,\cite{Huang08} which can deal, in principle, with large non-periodic systems have been formulated. In spite of all these developments, the calculation of the electronic structures of such systems remains computationally demanding, which makes the formulation of accessible approaches for this task a continuing effort in quantum chemistry.

Naturally, what makes such electronic-structure methods for non-periodic extended systems computationally expensive is their processing of information at the atomic (sub-nanometric) length scale. However, as the mesoscopic scale is approached, it can be expected that some physical (e.g. electronic and optical) properties of the system will depend on information at a larger (nanometric) scale. This means that these properties could be obtained from a course-grained average of the atomistic information. Hence, it is of considerable interest to devise methodologies that permit the direct determination, with good accuracy, of such properties bypassing the atomistic details. This would amount not only to large computational savings but also to increased understanding, since confounding atomistic information would be automatically averaged out.

An approach of this guise was introduced early in quantum chemistry for the determination of the electronic structure and spectra of linear\cite{Kuhn49} and cyclic\cite{Platt49} conjugated molecules, namely the `free-electron molecular-orbital' (FEMO) model. This approach, and extensions thereof,\cite{Ruedenberg53} were applied to small- and medium-size molecules. FEMO models have proven helpful as well in the study of other molecular properties, for instance the correlation of photoionization resonances with bond lengths\cite{Sheehy89} and the densities of   states in fullerenes.\cite{Mizorogi03}

An approach of the same spirit was also independently developed early in solid-state physics for the treatment of impurities and other slowly-varying weak perturbations in bulk semiconductors, namely the `envelope-function approximation' (EFA), also called `effective-mass approximation' (EMA).\cite{Wannier37,Luttinger55} Later, this method was successfully, albeit heuristically, applied to semiconductor quantum wells embedded in a bulk material.\cite{Dingle74} Since then, the EFA has provided a very useful framework for the interpretation and prediction of the electronic, optical and conduction properties of heterostructures built from crystalline materials,\cite{Yu99,Davies98,Jacak98} due to its relative conceptual simplicity and economical computational implementation in comparison with atomistic electronic-structure
 methodologies.\cite{Yu99,Davies98,DiCarlo03,Ladik99,Yu09,Hung09,Jacak98}

Although the EFA and the FEMO model have a lot in common, in particular the use of effective box potentials, the first has been put on a firm theoretical basis,\cite{Burt92,Foreman95} whereas the second is based on (very clever) \textit{ad hoc} considerations introduced on largely intuitive
grounds.

The main goal of this paper is to demonstrate that the formalism and implementation of the EFA can be straightforwardly, but rigorously, adapted for
the determination of the course-grained electronic structure of extended, albeit finite, molecular systems. This approach is hereafter called `envelope
molecular-orbital (EMO) theory'. Attention will be focused on organic quasi-linear heterostructures, for which it will become apparent that the
EMO method can readily handle systems with mesoscopic lengths.

Section \ref{sec:EMOTheory} begins with a compact review of the EFA formalism in the context of a polymer heterojunction, put in a language more quantum-chemical than found in the solid-state physics literature. The authors do not claim originality as to the formulation, which is standard, but
hope that their exposition will be helpful both to chemical physicists/quantum chemists, who may be unfamiliar with the subject, and solid-state
physicists interested in molecular nanostructures. Next, it is shown how the EFA can be rigorously adapted for (finite) molecules (EMO theory), by including explicitly the vacuum in the formalism. In Section \ref{sec:modelhetero} illustrative prototypes of quantum dots (QD's) are
designed by combining oligomers of \textit{trans}-polyacetylene (PA) and polydiacetylene (PDA). The EMO input parameters required
for such design are extracted from electronic-bandstructure calculations of these polymers in Section \ref{sec:implementation}. The results of the
EMO calculations for the prototypical QD's are presented, discussed and compared with the results of atomistic electronic-structure calculations in
Section \ref{sec:results}. Finally, Section \ref{sec:conclusions} provides concluding remarks and perspectives for future developments and
applications of the EMO theory.

\section{\label{sec:EMOTheory} Envelope Molecular-Orbital Theory}

\begin{figure}
\includegraphics[width=8.5cm]{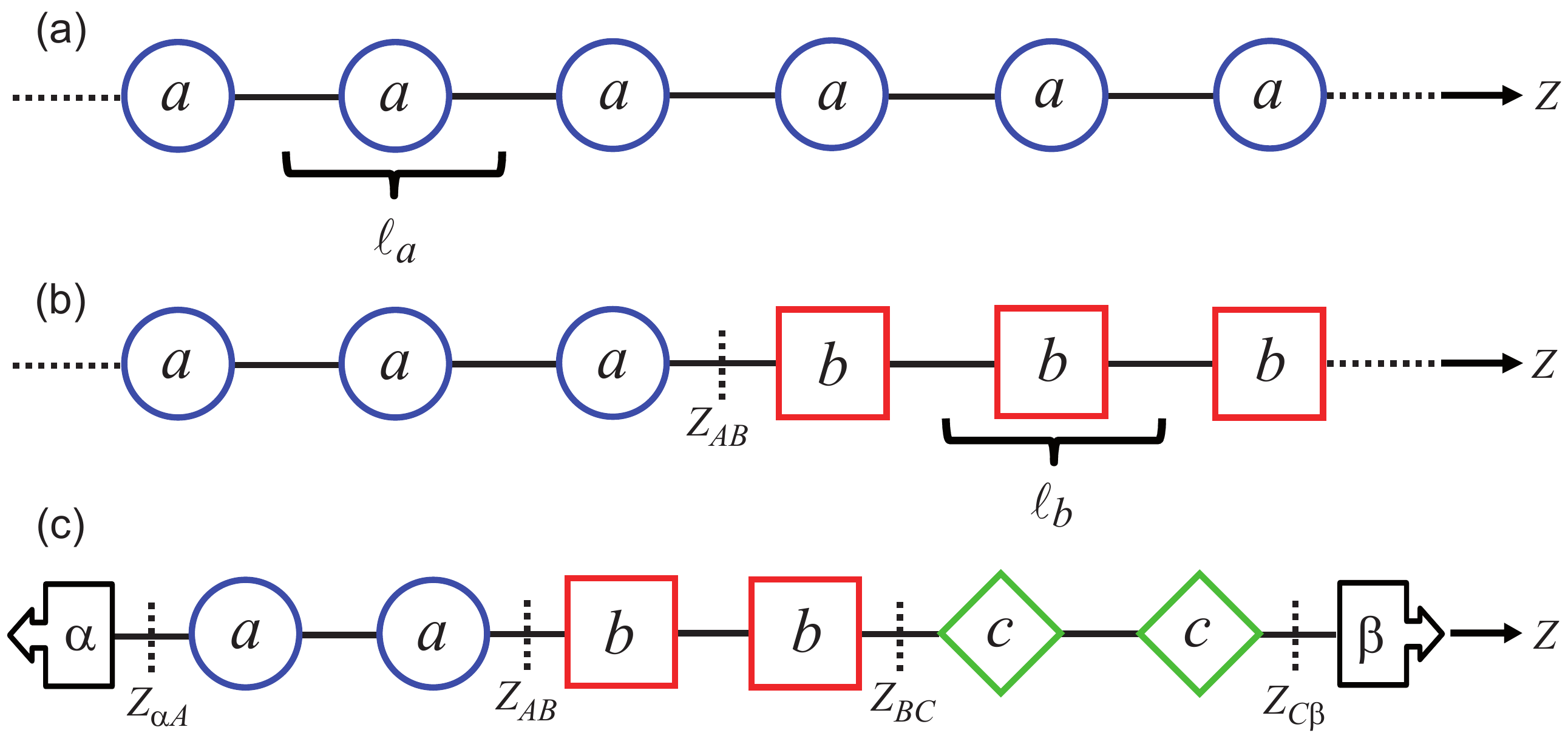}
\caption{\label{fig:1-lattices} (a) The infinite homopolymer $A$. (b) The infinite heteropolymer $A$-$B$. (c) The finite block co-oligomer $A$-$B$-$C$,
including the lateral vacua.}
\end{figure}

\begin{figure}
\includegraphics[width=8.5cm]{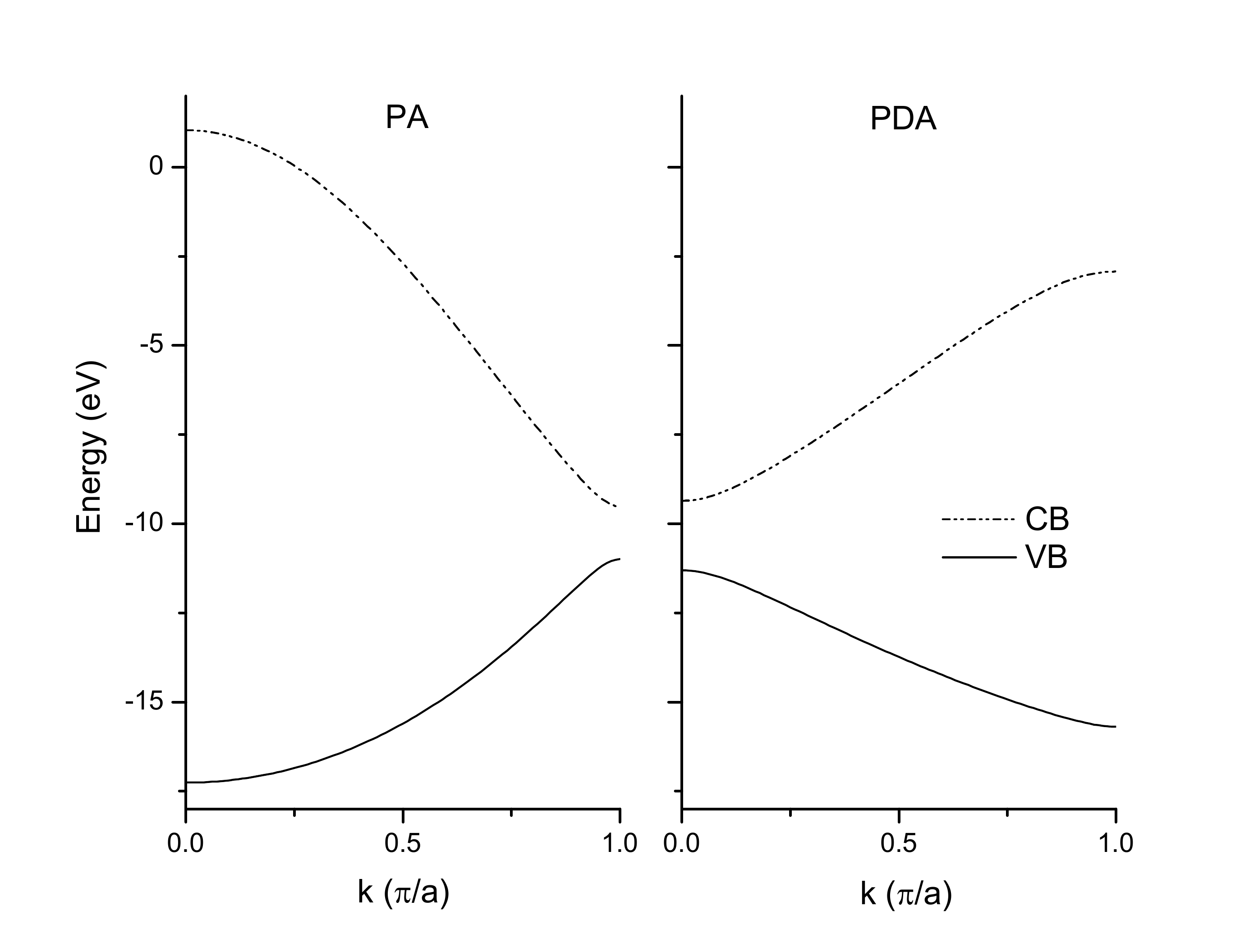}
\caption{\label{fig:2-bandsPA-PDA} EH bandstructures of the parent polymers. Due to $k/-k$ symmetry only half of the Brillouin zone is shown.}
\end{figure}

As a prelude, let us consider a $\pi$-conjugated polymer $A$ constituted of an infinite quasi-linear chain of identical monomers \textit{a} of length $\ell_a$ aligned in the $z$ direction (Fig. \ref{fig:1-lattices}(a)). As in any periodic system, the single-particle states, or crystalline
orbitals (CO's), of such polymer are organized into continuous bands.\cite{Ladik88,Hoffmann88} This work focuses on polymers with the highest
occupied ($\pi$) band, or valence band (VB), and the lowest unoccupied ($\pi^*$) band, or conduction band (CB), separated by an energy gap, i.e.
semiconductors or insulators. For example, the bandstructures of PA and PDA are displayed in Fig. \ref{fig:2-bandsPA-PDA}. According to Bloch's
theorem, each CO can be written as \cite{Ladik88,Hoffmann88}
\begin{equation}\label{eq:bloch}
\psi_n (k,\vec{r}) = e^{ikz} u_n (k,\vec{r}),
\end{equation}
where $n$ is the band index, $k \equiv k_z$ is the wavenumber conjugate to the longitudinal direction $z$, and $u_n (k,\vec{r})$ is the background Bloch function, which possesses the same spatial periodicity as the molecular potential. Since $\psi_n (k,\vec{r})$ is a periodic function of $k$ with period $2 \pi / \ell_a$, this variable is considered to be restricted to the first Brillouin zone (BZ), $|k| \leq \pi / \ell_a$. The CO's of polymers are commonly calculated by quantum-chemical methods, employing linear combinations of basis Bloch orbitals, which, in turn, are
translational-symmetry-adapted linear combinations of atomic orbitals (LCAO's).\cite{Ladik88,Hoffmann88,Ladik99}

Now, let us consider a $\pi$-conjugated heteropolymer $A$-$B$ formed by joining two semi-infinite strands of parent homopolymers $A$ and $B$ 
(Fig. \ref{fig:1-lattices}(b)). Due to the breaking of the spatial periodicity at the junction, Bloch's theorem (Eq. (\ref{eq:bloch})) no longer applies
and, consequently, the band structure of single-particle states is destroyed. Thus, these states must be described in terms of molecular orbitals
(MO's) instead of CO's. In a conventional quantum-chemical approach, the MO's are represented as LCAO's, or, if this system happens to belong to
one of the point-symmetry groups, as linear combinations of basis symmetry orbitals, which, in turn, are point-symmetry-adapted LCAO's.\cite{Levine,Schwartz04}
(In practice, since this hypothetical heteropolymer is infinite, artificial asymptotic or periodic boundary conditions must be imposed.)

Alternative approaches can be formulated, in which the MO's of the $A$-$B$ heteropolymer are constructed in terms of the CO's of the $A$ and $B$ parent homopolymers. For example, the Green's matrix formalism (GMF),\cite{Ladik87} has been developed for the atomistic determination of localized interface states. The EMO theory is in the same spirit, but it yields a non-atomistic (course-grained) approximation to the MO's.

The EMO formalism begins with a spatial partitioning of the single-particle Hamiltonian, as follows. Let us denote by $R_A$ and $R_B$ the spatial regions occupied by the semi-infinite strands of $A$ and $B$, and by $z_{AB}$ the $z$ coordinate of their junction, which can be regarded as the midpoint of the $A$-$B$ link (Fig. \ref{fig:1-lattices}(b)). In the $R_A$ region ($z < z_{AB}$) the Hamiltonian is expressed as
\begin{equation}\label{eq:partition}
\hat{H}^{(R_A)} = \hat{H}^{(R_A)}_A + U^{(R_A)}(\vec{r}),
\end{equation}
where $\hat{H}^{(R_A)}_A$ is the single-particle Hamiltonian of the parent homopolymer $A$ and $U^{(R_A)}(\vec{r})$ is a perturbing potential defined so as to take into account the \textit{changes} in the interactions acting on the particle in region $R_A$ due to the replacement of the semi-infinite strand of $A$ by the semi-infinite strand of $B$ in the $R_B$ region ($z > z_{AB}$). By definition, this potential is identically zero
for $z > z_{AB}$. For example, if $\hat{H}_A$ is the closed-shell Fock operator,\cite{Levine}
$\hat{H}_A \equiv \hat{f}_A = \hat{T} + \upsilon_{A,core}(\vec{r}) + \upsilon_{A,CX}(\vec{r})$, where the second and third terms are the core and Coulomb-plus-exchange
potentials, then
\begin{equation}\label{eq:pert}
\begin{split}
U^{(R_A)}(\vec{r}) &= [ \upsilon^{(R_B)}_{B,core}(\vec{r}) + \upsilon^{(R_B)}_{B,CX}(\vec{r}) - \upsilon^{(R_B)}_{A,core}(\vec{r})
\\ &  - \upsilon^{(R_B)}_{A,CX}(\vec{r}) ] \left[ 1 - \Theta(z - z_{AB}) \right],
\end{split}
\end{equation}
where $\Theta(z - z_{AB})$ is the Heaviside unit step function.\cite{Arfken} Although, for the present purposes, the precise form of the perturbing
potential need not be specified, it is evident that its magnitude is a maximum at the junction and usually small everywhere in comparison with the
magnitude of the effective potential present in $\hat{H}^{(R_A)}_A$ (in the example, $\upsilon^{(R_A)}_{A,core}(\vec{r}) + \upsilon^{(R_A)}_{A,CX}(\vec{r})$). Moreover, since the $\pi$ electron-nucleus and electron-electron interactions are screened, this potential can be expected to be of relatively short range. When a partitioning analogous to Eq. (\ref{eq:partition}) is defined for the $R_B$ region, the full single-particle Hamiltonian can be written in the spatially-piecewise fashion
\begin{equation}\label{eq:fullHam}
\hat{H} = \hat{H}^{(R_A)} \left[ 1 - \Theta(z - z_{AB}) \right] + \hat{H}^{(R_B)} \Theta(z - z_{AB}).
\end{equation}

The CO's of the parent polymers $J=A,B$ are the eigenfunctions of respective single-particle Hamiltonians,
\begin{equation}\label{eq:SE0}
\hat{H}_{J} \psi^{(J)}_n (k,\vec{r}) = \varepsilon^{(J)}_n (k) \psi^{(J)}_n (k,\vec{r}),
\end{equation}
and, consequently, constitute a complete orthonormal set,
\begin{equation}\label{eq:orthopsi}
\int_{R^3} d^3 r \psi^{(J)*}_m (k',\vec{r}) \psi^{(J)}_n (k,\vec{r}) = \delta (k'-k) \delta_{mn}.
\end{equation}
Therefore, the MO's of the $A$-$B$ heteropolymer, which are the eigenfunctions of the single-particle Hamiltonian (\ref{eq:fullHam}),
\begin{equation}\label{eq:SE}
\hat{H} \Psi(\vec{r}) = E \Psi(\vec{r}),
\end{equation}
can be represented \textit{in each region} in terms of the CO's of the corresponding parent homopolymer. In the $R_J$ region such `linear
combination of crystalline orbitals' (LCCO) reads
%
%
\begin{equation}\label{eq:LCCO}
\begin{split}
\Psi^{(R_J)} (\vec{r}) &= \sum_{n} \int_{BZ} dk f^{(R_J)}_n(k) \psi^{(J)}_n(k,\vec{r})  \\&= \sum_{n} \int_{BZ} dk e^{ikz} f^{(R_J)}_n(k) u^{(J)}_n(k,\vec{r}),
\end{split}
\end{equation}
%
where the amplitudes $f^{(R_J)}_n(k)$ are assumed to possess the same $k$-periodicity as the CO's and Bloch's theorem (Eq. (\ref{eq:bloch})) was used
in the second equation. Substitution of this expansion into Eq. (\ref{eq:SE}) followed by projection onto $\psi^{(J)}_m (k',\vec{r})$ and Fourier transformation to $z$-space yield the basic EMO equations
%
%
\begin{equation}\label{eq:EFE}
\begin{split}
\hat{\varepsilon}^{(J)}_m (\hat{K}) F^{(R_J)}_m (z) &+ \sum_{n} \int_{R} dz' U^{(R_J)}_{mn} (z,z') F^{(R_J)}_n (z') \\& = E F^{(R_J)}_m (z),
\end{split}
\end{equation}
%
where Eqs. (\ref{eq:fullHam}), (\ref{eq:partition}), (\ref{eq:SE0}) and (\ref{eq:orthopsi}) were taken into account, $\hat{\varepsilon}^{(J)}_m (\hat{K})$ is a `generalized kinetic-energy operator' obtained by formally replacing the variable $k$ by the operator $\hat{K} = -i\frac{d}{dz}$ in the dispersion relation of Eq. (\ref{eq:SE0}), 
%
\begin{equation}\label{eq:Umn}
\begin{split}
U^{(R_J)}_{mn} (z,z')  & \equiv \dfrac{\ell_j}{2 \pi} \int_{BZ} dk \int_{BZ} dk' e^{ik'z}  e^{-ikz'} \\& \times\left[ \int_{R^3} d^3 r \psi^{(J)*}_m(k',\vec{r}) U^{(R_J)}(\vec{r}) \psi^{(J)}_m(k,\vec{r}) \right],
\end{split}
\end{equation}
%
and $F^{(R_J)}_m (z)$ is an `envelope MO' defined by
%
\begin{equation}\label{eq:defenv}
F^{(R_J)}_m (z) \equiv \dfrac{\ell_j}{2 \pi} \int_{BZ} dk e^{ikz} f^{(R_J)}_m (k).
\end{equation}
%
%
%
%
with $j=a,b$. (It should be noticed that the functions $\Psi^{R_J}(\vec{r})$ and $F^{R_J}_m(z)$ are defined over the entire $z-$space, the super-index indicating that only their values confined within the $R_J$ region are being considered.) In obtaining the first term of Eq. (\ref{eq:EFE}) it was recognized that
%
%
\bigskip
\bigskip
\bigskip
\bigskip
\bigskip
\begin{widetext}
\begin{eqnarray}\label{eq:genkin} \nonumber
\dfrac{\ell_j}{2 \pi} \int_{BZ} dk e^{ikz} \varepsilon^{(J)}_m(k) f^{(R_J)}_m(k) & = & \int_{R} dz' \tilde{\varepsilon}^{(J)}_m(z') F^{(R_J)}_m(z-z') \\ \nonumber
& = & \left( \int_{R} dz' \tilde{\varepsilon}^{(J)}_m(z') e^{-iz'\hat{K}} \right) F^{(R_J)}_m(z) \\
& = & \hat{\varepsilon}^{(J)}_m (\hat{K}) F^{(R_J)}_m (z),
\end{eqnarray}
\end{widetext}
where in the first line the convolution theorem\cite{Arfken} was used, with
\begin{equation}\label{eq:FTdisp}
\tilde{\varepsilon}^{(J)}_m(z) \equiv \dfrac{\ell_j}{2 \pi} \int_{BZ} dk e^{ikz} \varepsilon^{(J)}_m(k),
\end{equation}
in the second line the translation operator\cite{Yu99} $\hat{\tau}(z') F^{(R_J)}_m(z) \equiv \exp(-iz'\hat{K}) F^{(R_J)}_m(z) = F^{(R_J)}_m(z-z')$
was introduced, and in the third line the inverse of Eq. (\ref{eq:FTdisp}) was used formally replacing the variable $k$ by the
operator $\hat{K} = -i\frac{d}{dz}$.

The set of real-space integro-differential equations (\ref{eq:EFE}) for the envelope MO's is equivalent to the Schr\"odinger equation (\ref{eq:SE}), but
much more difficult to solve in practice. Sophisticated techniques for dealing with the nonlocal terms $U^{(R_J)}_{mn} (z,z')$ have been
developed.\cite{Foreman95} Fortunately, these equations are also amenable to approximations that simplify drastically the determination of the MO's,
at the same time retaining the essential features necessary for describing the physics of the system. Since the main purpose of this paper is
to show that Eqs. (\ref{eq:EFE}) can be fruitfully adapted to molecules, the second, much simpler approach is adopted here.

The \textit{first} and key
approximation, which will be called `uncoupled approximation', consists of neglecting altogether the nonlocal perturbation matrix
elements $U^{(R_J)}_{mn} (z,z')$. This approximation is valid sufficiently far from the heterojunction, where $U^{(R_J)} (\vec{r})$ is not large enough to couple the bands appreciably. The resulting uncoupled equations for the envelope MO's are free-particle-like Schr\"odinger equations with generalized kinetic-energy operators $\hat{\varepsilon}^{(J)}_m (\hat{K})$. Now, in both regions each MO (Eq. \ref{eq:LCCO}) contains contributions from one band only, for which reason this approximation is also termed `one-band approximation'. It is reasonable to expect that the interfacial region around the junction, where this approximation fails, has a width $w_{AB}\sim\ell_a + \ell_b$.

For the sake of concreteness, from now on interest will be focused on occupied (hole) MO's with energies near the VB edges (or highest occupied
crystalline orbitals (HOCO's)), $\varepsilon^{(J)}_v (k = k_v)$, and virtual (electron) MO's with energies near the CB edges (or lowest unoccupied
crystalline orbitals (LUCO's)), $\varepsilon^{(J)}_c (k = k_c)$, of the parent polymers, since their optical and conduction properties are largely dominated by these $\pi$ and $\pi^*$ states, respectively.\cite{Ladik88} Furthermore, let us assume for the time being that the HOCO and
LUCO of homopolymer $A$ lie higher and lower, respectively, than the HOCO and LUCO of homopolymer $B$ (see Fig. (\ref{fig:2-bandsPA-PDA})).

In general, three types of MO are possible in such $A$-$B$ heteropolymer: (1) For holes with
$\varepsilon^{(B)}_v (k_v) \leq E \leq \varepsilon^{(A)}_v (k_v)$ or electrons with
$\varepsilon^{(A)}_c (k_c) \leq E \leq \varepsilon^{(B)}_c (k_c)$, MO's confined mainly to the (classically-allowed) $R_A$ region and
decaying into the (classically-forbidden) $R_B$ region; (2) for holes with $E < \varepsilon^{(B)}_v (k_v)$ or electrons
with $E > \varepsilon^{(B)}_c (k_c)$, MO's delocalized over the entire structure; and (3) MO's localized around the junction (``interface''
states), which, evidently, cannot be treated within the uncoupled approximation and will not be considered in this work. MO's confined to the $R_B$ region and decaying into the $R_A$ region are impossible with the band alignment presently assumed.

Let us consider a type (1) MO with energy $E$. In the $R_A$ region, and outside the interface, such MO must be very similar to the CO of parent
homopolymer $A$ with energy $E = \varepsilon^{(A)}_n (k = k_E)$, which implies that the function $f^{(R_A)}_n (k)$ must be highly peaked around
$k_E$. Hence, since $u^{(A)}_n (k,\vec{r})$ is a slowly-varying function of $k$, it is a good approximation to neglect its $k$ dependence  within the range of $f^{(R_A)}_n (k)$ and take it out of the integral in Eq. (\ref{eq:LCCO}). On the other hand, in the $R_B$ region at energy $E$ there are not any states of the parent homopolymer $B$, which implies that the MO must comprise contributions from CO's of $B$ with energies spanning, in principle, the entire BZ, in order to achieve the destructive interferences necessary for the formation of an evanescent wave. Thus, the function $f^{(R_B)}_n (k)$ must be much wider around $k_E$ than $f^{(R_A)}_n (k)$ and the neglect of the $k$ dependence of $u^{(B)}_n (k,\vec{r})$ in Eq. (\ref{eq:LCCO}) is unjustified. Nevertheless, since the evanescent portion of the MO is of short range, the relative error introduced by this approximation in the $R_B$ region should be actually small. By the same token, outside the interface a MO of type (2) with energy $E$ resembles the CO's of parent homopolymers $A$ and $B$ with energies $E = \varepsilon^{(A)}_n (k_E)$ and $E = \varepsilon^{(B)}_n (k_E)$ in the $R_A$ and $R_B$ regions, respectively. In conclusion, taking into account Eq. (\ref{eq:defenv}) and the assumed proximity to the respective band edges, in both regions hole and electron MO's of types (1) and (2) are represented as
%
\begin{eqnarray}\label{eq:MO-EFA}
\label{eq:MOEFAv}
 \Psi^{(R_J)}_h(\vec{r}) & \cong & F^{(R_J)}_h(z) u^{(J)}_v(k_v,\vec{r}), \\
\label{eq:MOEFAc}
 \Psi^{(R_J)}_e(\vec{r}) & \cong & F^{(R_J)}_e(z) u^{(J)}_c(k_c,\vec{r}).
\end{eqnarray}
%
These expressions constitute the \textit{second} approximation of the development and will be termed ``single-envelope MO approximation".

At this point, it is important to emphasize that, since $f^{(R_A)}_n (k)$ is highly peaked around $k_n$, according to Eq. (\ref{eq:defenv})
$F^{(R_A)}_n (z)$ must be a slowly-varying function of $z$. This can be seen by considering the limit $f^{(R_A)}_n(k) \rightarrow \delta (k-k_E)$, where $F^{(R_A)}_n (z)$ becomes a plane wave and Bloch's theorem (\ref{eq:bloch}) is recovered in Eq. (\ref{eq:LCCO}): since $\abs{k_{max}} = \pi / \ell_a = 2\pi / \lambda_{min}$, then $\lambda_{min} = 2 \ell_a$, indicating that the EMO's vary appreciably only over a range of several monomers, in contrast with $u^{(A)}_n (k,\vec{r})$ which varies widely from one monomer to the next. Thus, Eqs. (\ref{eq:MOEFAv}) and (\ref{eq:MOEFAc}) indicate that an MO is approximately constituted by a (rapidly-varying) background Bloch function modulated by a slowly-varying
EMO, which is analogous to Bloch's theorem for a CO (Eq. (\ref{eq:bloch})) where a background Bloch function is modulated by a
slowly-varying plane wave.

Let us now examine more closely the nature of the operator $\varepsilon^{(J)}_n (\hat{K})$. Near the HOCO and LUCO, $k_n$ ($n=v,c$), where the
energy dispersion is usually nearly parabolic (see Fig. \ref{fig:2-bandsPA-PDA}), this function can be expressed as
$\varepsilon^{(J)}_n (k - k_n) \cong \varepsilon^{(J)}_n (k = k_n) + \left( \frac{d^2 \varepsilon^{(J)}_n}{dk^2} \right)_{k=k_n} \frac{(k - k_n)^2}{2}$.
In order to make an analogy with the free-particle dispersion, the `effective mass'
\begin{equation}\label{eq:defPEM}
\dfrac{1}{m^{(J)*}_n} \equiv \dfrac{1}{\hslash^2} \left( \dfrac{d^2 \varepsilon^{(J)}_n}{dk^2} \right)_{k=k_n}
\end{equation}
is introduced. (It should be noticed that, since the VB dispersion has the shape of an inverted parabola around the HOCO, the effective mass of the
holes is negative.) Formal replacement of the variable $k - k_n$ by the operator $\hat{K} = -i\frac{d}{dz}$ yields
%
\begin{eqnarray}\label{eq:defPGKEO}
\nonumber\hat{\varepsilon}^{(J)}_n (\hat{K}) & \cong & \varepsilon^{(J)}_n (k = k_n) + \dfrac{\hslash^2 \hat{K}^2}{2 m^{(J)*}_n} \\
& = & \varepsilon^{(J)}_n (k = k_n) - \dfrac{\hslash^2}{2 m^{(J)*}_n} \dfrac{d^2}{dz^2}
\end{eqnarray}
%
In this expression, $k_n=0$ or $k_n=\pi/\ell_a$, depending on how the VB and CB of the $J$ homopolymer run.\cite{Ladik88,Hoffmann88} For example, in
Fig. \ref{fig:2-bandsPA-PDA} it can be seen that the frontier CO's of PA are located at the edge of the BZ, whereas the frontier CO's of PDA are
located at the center of the BZ. Eq. (\ref{eq:defPGKEO}) constitutes the \textit{third} approximation of the development, the familiar `parabolic
EMA', \cite{Yu99,Davies98,Burt92} whence $\hat{\varepsilon}^{(J)}_n (\hat{K})$ becomes the standard kinetic-energy operator, with the
energy measured from the band extremum. Now, Eq. (\ref{eq:EFE}) reduces to the free-particle Schr\"odinger equation
%
\begin{equation}\label{eq:EM-SE}
\left( - \dfrac{\hslash^2}{2 m^{(J)*}_n} \dfrac{d^2}{dz^2} + \varepsilon^{(J)}_n (k_n) \right) F^{(R_J)}_n (z) \cong E F^{(R_J)}_n (z).
\end{equation}
%

When both regions are considered together the following picture emerges: To the left and right of the interface the carriers
behave like free particles with effective masses $m^{(A)*}_n$ and $m^{(B)*}_n$ subjected to constant potentials $\varepsilon^{(A)}_n (k_n)$ and $\varepsilon^{(B)}_n (k_n)$, respectively. However, within
the interface, where Eqs. (\ref{eq:MOEFAv}), (\ref{eq:MOEFAc}) and (\ref{eq:EM-SE}) do not hold, the behavior of these quasi-particles remains unknown. In order to obtain a well-posed Sturm-Liouville problem for the EMO's across the entire structure, Eq. (\ref{eq:EM-SE}) is artificially extended into the interface. This \textit{fourth} approximation is reasonable because, as discussed above, EMO's of types (1) and (2) are expected to vary little across the interfacial region. This extension can be accomplished by introducing any smooth interpolations across $w_{AB}$ of the band-extremum potentials and effective masses between the pairs of values $\varepsilon^{(A)}_n (k_n)$,$\varepsilon^{(B)}_n (k_n)$ and $m^{(A)*}_n$,$m^{(B)}_n$, respectively. 
Now, Eqs. (\ref{eq:EM-SE}) for both regions can be comprised into the single equation valid across the entire structure
%
\begin{equation}\label{eq:fullEM-SE}
\left[ - \dfrac{\hslash^2}{2} \dfrac{d}{dz} \left( \dfrac{1}{m^*_n (z)} \dfrac{d}{dz} \right)  + V_n(z) \right] F_n(z) \cong E F_n(z),
\end{equation}
%
where $m^*_n (z)$ and $V_n(z)$ are the position-dependent effective mass and band-extremum potential, respectively, and the BenDaniel-Duke expression
for the kinetic-energy operator with a position-dependent mass\cite{BenDaniel66} has been used.

This procedure automatically enforces the continuity of the EMO's and their derivatives at the heterojunction.\cite{BenDaniel66} However, it should be mentioned that, according to Eqs. (\ref{eq:MOEFAv}) and (\ref{eq:MOEFAc}), the continuity of an EMO apparently does not guarantee the continuity of the full MO at the heterojunction, since $u^{(A)}_n(k_n,\vec{r})$ and $u^{(B)}_n(k_n,\vec{r})$ do not, in general, match at that point.\cite{Burt92} Nevertheless, it can be shown that this is not the case, by taking into account the phase invariance of the Bloch orbitals (Eq. \ref{eq:bloch}).\cite{Arce}

Finally, let us consider a \textit{finite} $n$-block $\pi$-conjugated co-oligomer $A$-$B$-$\cdots$-$Z$ formed by joining a series of $n$ oligomers of parent polymers $A$, $B$, $\cdots Z$. For example, Fig. \ref{fig:1-lattices}(c) illustrates a three-block structure $A$-$B$-$C$. In general, three types of bound MO's are possible in this system: (1) MO's confined mainly to one or several oligomers
and decaying into the other regions; (2) MO's delocalized over the entire structure; (3) MO's localized around a junction. Specific examples of
type (1) and (2) MO's are found in Section \ref{sec:results}.

The EMO formalism for a single heterojunction can be extended to such molecules by making the following considerations: (1) The
key idea is to regard the co-oligomer as an $\alpha/A/B/\cdots/Y/Z/\beta$ heterostructure, where $\alpha$ and $\beta$ stand for the semi-infinite
left and right vacua, respectively (Fig. \ref{fig:1-lattices}(c)). (2) To accommodate these vacua in the development, they are considered as
hypothetical parent polymers endowed with empty lattices. (3) Now, the left and right chain ends can be formally regarded as $\alpha-A$ and
$Z-\beta$ heterojunctions, respectively. (4) Finally, each heterojunction is treated as if it were isolated from the rest.

Assumption (4) is justified on the following grounds. The perturbing potential in a given region (oligomer or vacuum) now contains contributions from the effective single-particle potentials of all the other regions. However, only the nearest-neighboring regions contribute appreciably to this potential (two to each oligomer and one to each vacuum), since the others are too far away. The full single-particle Hamiltonian now assumes the spatially-piecewise form
%
\begin{eqnarray}\label{eq:fullHam2}
\hat{H} & = & \hat{H}^{(R_{\alpha})} \left[ 1 - \Theta(z - z_{\alpha A}) \right] \\ \nonumber
& + & \sum_{i=A}^{Y} \hat{H}^{(R_i)} \left[ \Theta(z - z_{i-1,i}) - \Theta(z - z_{i,i+1}) \right] \\ \nonumber
& + & \hat{H}^{(R_{\beta})} \Theta(z - z_{Z\beta}).
\end{eqnarray}
%
Thus, the uncoupled approximation can still be used for all the oligomer-oligomer heterojunctions. On the other hand, this approximation breaks down at the vacuum-oligomer interfaces, for the following reason: The single-particle Hamiltonian in a vacuum region of the
heterostructure, say $R_{\alpha}$, has the form $\hat{H}^{(R_{\alpha})} = \hat{T} + U^{(R_{\alpha})}(\vec{r})$. In this case the
magnitude of the perturbing potential cannot be considered either small or short-ranged, since the effective potential in the Hamiltonian of
the ``polymer'' $\alpha$ is null. For example, if the effective potential of the $A$ oligomer is the closed-shell
Coulomb-plus-exchange potential, then $U^{(R_\alpha)}(\vec{r}) = \left[ \upsilon^{(R_A)}_{A,core}(\vec{r}) +
\upsilon^{(R_A)}_{A,CX}(\vec{r}) \right] \left[ 1 - \Theta(z - z_{\alpha A}) \right]$. By the same token, the contribution of $\alpha$
to the perturbing potential in the $R_A$ region would be $\left[ - \upsilon^{(R_{\alpha})}_{A,core}(\vec{r}) -
\upsilon^{(R_{\alpha})}_{A,CX}(\vec{r}) \right] \Theta(z - z_{\alpha A})$, which is not small either. Nevertheless, since the $\alpha$-$A$ and $Z$-$\beta$ interfaces are short in comparison with the length of the molecule, and an MO whose amplitude reaches a chain end must decay rapidly as it penetrates into the vacuum, this breakdown is expected to cause only small errors in practice.

Consideration (2) means that each vacuum ``polymer'' possesses a parabolic (free-particle) dispersion relation around any $k\equiv k_z$ and
CO's with the forms $\psi^{(\gamma)}_n (k,\vec{r}) = e^{ikz} u_n^{(\gamma)} (x,y)$, where $u_n^{(\gamma)}(x,y)= e^{ik_xx}e^{ik_yy}$ with $\gamma=\alpha,\beta$ (see Eq. (\ref{eq:bloch})). Therefore, the single-envelope MO approximation (Eqs. (\ref{eq:MOEFAv}) and (\ref{eq:MOEFAc})) and the parabolic EMA (Eq. (\ref{eq:defPGKEO})) are actually exact in the $R_{\alpha}$ and $R_{\beta}$ regions. Moreover, within the molecule the single-envelope approximation continues to be valid outside the interfacial regions, and, for a given MO, the parabolic approximation works best in regions where its energy is close to the corresponding parent polymer band extremum.

Now, the hole and electron EMO's are determined across the entire molecule by Eq. (\ref{eq:fullEM-SE}), where the potential-energy functions and position-dependent effective masses assume one-dimensional piecewise-constant profiles, except within the interfaces where smooth interpolations of these functions are employed.

\begin{figure}
\includegraphics[width=8.5cm]{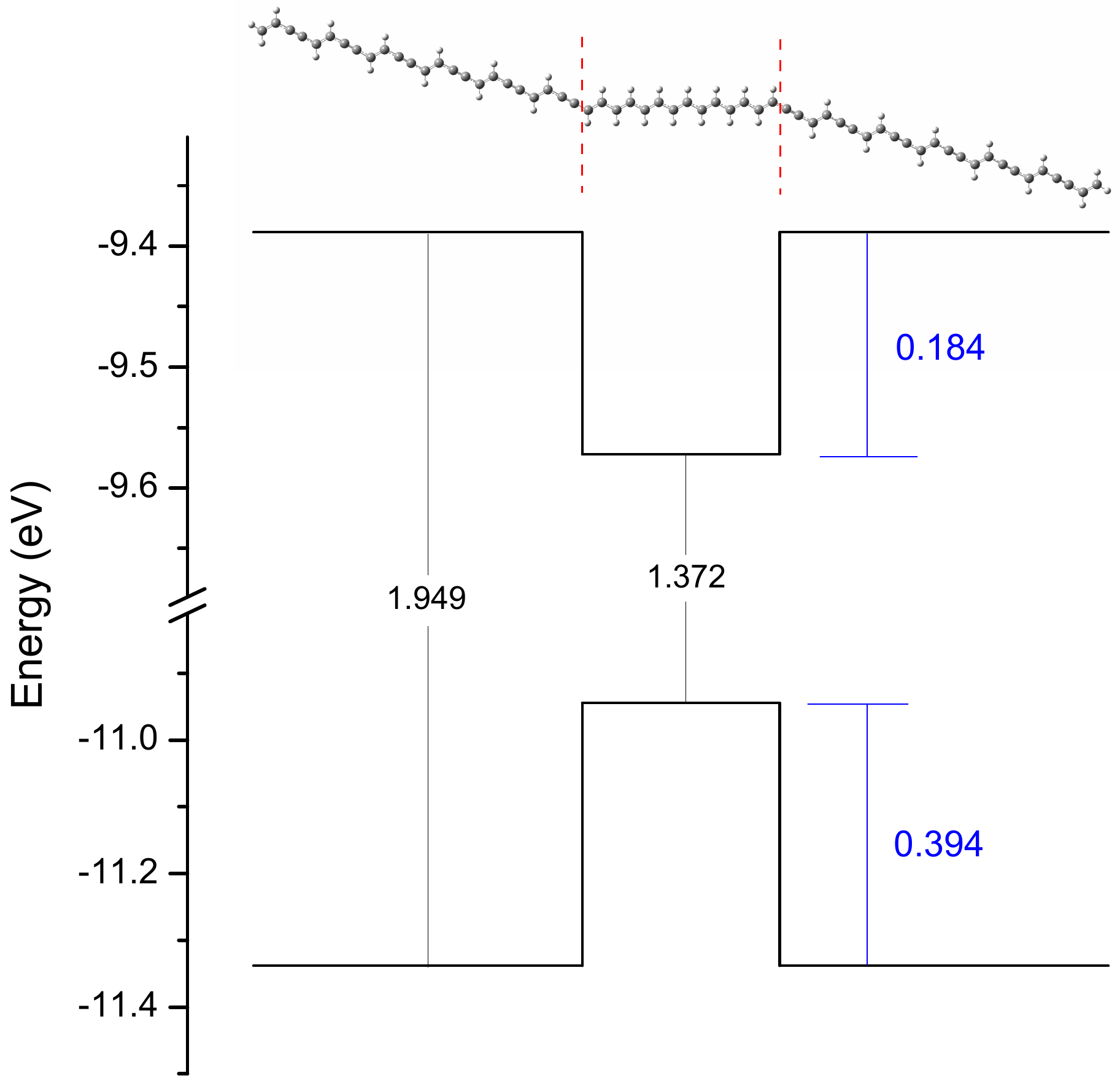}
\caption{\label{fig:3_scheme-PDA-PA-PDA} Structure and intramolecular electron and hole potential-energy profiles, according to the EH alignments, of the co-oligomer PDA$_6$/PA$_7$/PDA$_6$. The heterojunctions are drawn as abrupt and the vacuum and LOCO levels are not shown.}
\end{figure}

Fig. \ref{fig:3_scheme-PDA-PA-PDA} illustrates an example of a three-block co-oligomer of the $A$-$B$-$A$ kind, together with the piecewise-constant intramolecular potential profiles for electrons and holes. An intramolecular barrier height for electrons or holes is given by the difference between the energies (`offset') of the LUCO's or the HOCO's, respectively, of the neighboring parent polymers,
\begin{equation}\label{eq:offset}
V_n \equiv |\varepsilon^{(PA)}_n (k_n) - \varepsilon^{(PDA)}_n (k_n)|.
\end{equation}
By the same token, a molecule-vacuum barrier height for electrons is given by the offset between the LUCO of the lateral parent polymer (in the example, PDA) and the LUCO of the vacuum (which lies at zero energy). On the other hand, a molecule-vacuum barrier height for holes cannot be defined in an analogous way because the vacuum does not possess a HOCO. Nevertheless, the molecule-vacuum junction for holes can be handled by taking into account that the hole spectrum of the lateral parent polymer is bounded from above and from below by the energies of its HOCO and lowest \textit{occupied} crystalline orbital (LOCO), respectively, and that the hole effective mass is negative. The molecule-vacuum junctions are not indicated in Fig. \ref{fig:3_scheme-PDA-PA-PDA}, since the vacuum and LOCO levels lie very high up and deep down, respectively. Naturally, the magnitudes of the electron and hole effective masses in the vacuum regions are equal to the normal electron mass, $m_0$.

\section{\label{sec:modelhetero} Model Heterostructures}

\begin{figure*}
\includegraphics[width=10.0cm]{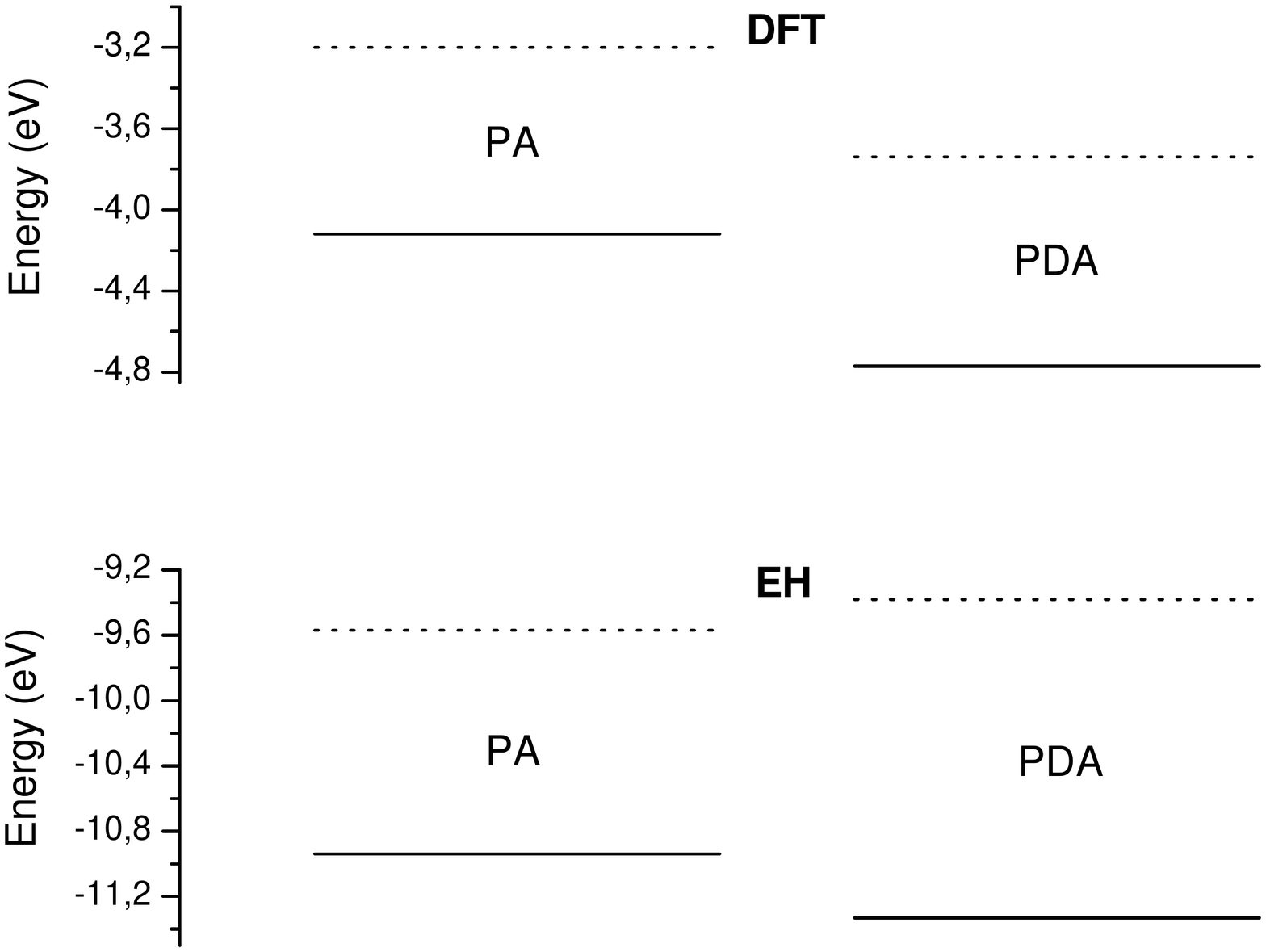}
\caption{\label{fig:4-band align} DFT and EH band alignments of the parent polymers. Dotted lines: LUCO's, solid lines: HOCO's.}
\end{figure*}

To illustrate the theory, simple prototypical $n$-block ($n$=1,2,3) structures were studied, employing the common parent polymers PA and PDA. These are shown in the first columns of Tables \ref{tab:table I} and \ref{tab:table II}, where the subscripts indicate the numbers of monomers in the constituent oligomers and the vacua have not been indicated. The EMO intramolecular potential profiles for electrons and holes in these systems are determined by the alignments of the energies of the LUCO's and the HOCO's, respectively, of PA and PDA. In Section \ref{sec:implementation}, they are obtained by means of two methods and the results are presented in Tables \ref{tab:table I} and \ref{tab:table II} and Fig. \ref{fig:4-band align}. On the other hand, the LOCO energies, required for handling the oligomer-vacuum junction for holes, are not available. However, in practice this does not pose a problem, since the energy of this state is so far down that its magnitude can be taken as infinite without incurring in any significant errors.

Fig. \ref{fig:3_scheme-PDA-PA-PDA} shows the resulting intramolecular profiles for a symmetrical three-block structure. For interpretation purposes, it is useful to notice that although the intramolecular profile for holes looks like a potential barrier, it can be thought of as a well if the effective mass of the holes is taken as positive.

These quasi-linear architectures constitute QD's\cite{Yu99,Davies98,Jacak98,note1} since the particles can be confined in the three directions. The design of molecules of this type can be useful for optical and electronic applications,\cite{Bredas90} since they can exhibit relatively localized states with controllable sizes and energies, as will be appreciated in Section \ref{sec:results}.

Studies of the electronic states of heterostructures similar to these have been reported by other authors, employing conventional quantum-chemical or solid-state methods.\cite{Ruckh86, Bakhshi92, Meyers92, Liu00, Mingo01, Cheng07}

\begin{table*}
\caption{\label{tab:table I} EH values of the effective masses (in units of the vacuum electron mass, $m_0$) and the frontier CO energies of the parent polymers, and the high-lying hole and low-lying electron energies of the $n$-block structures. A state marked with an asterisk is unbound with respect to the corresponding intramolecular well. $\delta E \equiv |E_{EH}-E_{EMO}|$.}
\begin{ruledtabular}
\begin{tabular}{ccccccc}
              System                   &      $m^*_h$    &     $m^*_e$     &   State     &    EH    &   EMO    &      $\delta E$      \\
                                       &      ($m_0$)    &     ($m_0$)     &             &   (eV)   &   (eV)   &       (meV/mEh)      \\
\hline
                PA                     &      0.041      &      0.037      &   HOCO      & -10.944  &          &                      \\
                                       &                 &                 &   LUCO      &  -9.572  &          &                      \\
\hline
                PDA                    &      0.048      &      0.034      &   HOCO      & -11.338  &          &                      \\
                                       &                 &                 &   LUCO      &  -9.389  &          &                      \\
\hline
                                       &                 &                 &   HOMO-3    & -11.243  & -11.304  &         61/2.2       \\
                                       &                 &                 &   HOMO-2    & -11.129  & -11.158  &         29/1.1       \\
                                       &                 &                 &   HOMO-1    & -11.033  & -11.041  &         8/0.29       \\
PDA$_{52}$/PA$_{65}$/PDA$_{52}$        &                 &                 &   HOMO      & -10.968  & -10.969  &         1/0.037      \\
                                       &                 &                 &   LUMO      &  -9.552  &  -9.548  &         4/0.15       \\
                                       &                 &                 &   LUMO+1    &  -9.495  &  -9.480  &         15/0.55      \\
                                       &                 &                 &   LUMO+2    &  -9.414  &  -9.393  &         21/0.77      \\
                                       &                 &                 &   LUMO+3    &  -9.379$^*$ &  -9.373$^*$ &   6/0.22       \\
\hline
                                       &                 &                 &   HOMO-2    & -11.294  & -11.340$^*$ &   46/1.7       \\
                                       &                 &                 &   HOMO-1    & -11.128  & -11.146  &         18/0.66      \\
PDA$_{26}$/PA$_{40}$/PDA$_{26}$        &                 &                 &   HOMO      & -10.995  & -10.997  &         2/0.073      \\
                                       &                 &                 &   LUMO      &  -9.532  &  -9.524  &         8/0.29       \\
                                       &                 &                 &   LUMO+1    &  -9.428  &  -9.409  &         19/0.70      \\
                                       &                 &                 &   LUMO+2    &  -9.359$^*$ &  -9.338$^*$ &   21/0.77      \\
\hline                                 
                                       &                 &                 &   HOMO-1    & -11.379$^*$  & -11.412$^*$ &  33/1.21      \\
      PDA$_{20}$/PA$_{20}$             &                 &                 &   HOMO      & -11.143  & -11.152  &         9/0.331      \\
                                       &                 &                 &   LUMO      &  -9.411  &  -9.410  &         1/0.0367     \\      
                                       &                 &                 &   LUMO+1    &  -9.320$^*$  &  -9.248$^*$ &   72/2.65      \\
\hline  
                                      &                 &                 &   HOMO-2     & -11.203  &  -11.475 &          272/9.99    \\ 
                                      &                 &                 &   HOMO-1     & -11.093  &  -11.089 &          4/0.146     \\ 
          PA$_{65}$                   &                 &                 &   HOMO       & -11.020  &  -10.980 &          40/1.47     \\ 
                                      &                 &                 &   LUMO       & -9.491   &  -9.531  &          40/1.47     \\
                                      &                 &                 &   LUMO+1     & -9.416   &  -9.409  &          7/0.257     \\
                                      &                 &                 &   LUMO+2     & -9.301   &   -9.207 &          91/3.34     \\                              
\end{tabular}
\end{ruledtabular}
\end{table*}

\begin{table*}
\caption{\label{tab:table II} Analogous to Table I, employing DFT instead of EH.}
\begin{ruledtabular}
\begin{tabular}{ccccccc}
              System                   &      $m^*_h$    &     $m^*_e$     &   State     &   DFT    &   EMO    &      $\delta E$      \\
                                       &      ($m_0$)    &     ($m_0$)     &             &   (eV)   &   (eV)   &       (meV/mEh)      \\
\hline
                PA                     &     0.066      &      0.063     &   HOCO        &  -4.120  &          &                      \\
                                       &                 &               &   LUCO        &  -3.207  &          &                      \\
\hline 
                PDA                    &     0.075      &      0.065      &   HOCO       &  -4.771  &          &                      \\
                                       &                 &                &   LUCO       & -3.740   &          &                      \\
\hline
                                       &                 &                 &   HOMO-2    &  -4.656  & -4.834$^*$   &         180/6.6    \\
                                       &                 &                 &   HOMO-1    &  -4.515  &	-4.708	  &         190/7.0      \\
       PDA$_{20}$/PA$_{20}$            &                 &                 &   HOMO      & -4.260	  & -4.283    &         23/0.85      \\
                                       &                 &                 &   LUMO      & -3.582   &	-3.689	  &         110/4.0    \\
                                       &                 &                 &   LUMO+1    & -3.459	  & -3.541	  &         82/3.0     \\
                                       &                 &                 &   LUMO+2    & -3.317   & -3.305    &         12/0.44   \\
                                       &                 &                 &   LUMO+3    & -3.186$^*$  & -3.057$^*$   &    129/4.7   \\
\hline 
                                       &                 &                 &   HOMO-2     & -4.982   & -5.247   &         265/9.73     \\ 
                                       &                 &                 &   HOMO-1     & -4.835   & -4.983   &         148/5.44     \\ 
              PDA$_{20}$               &                 &                 &   HOMO       & -4.735   & -4.824   &          89/3.3     \\ 
                                       &                 &                 &   LUMO       & -3.656   & -3.679   &         23/0.84     \\
                                       &                 &                 &   LUMO+1     & -3.541   & -3.496   &         45/1.65    \\
                                       &                 &                 &   LUMO+2     & -3.378   & -3.191   &         187/6.87   \\

\end{tabular}
\end{ruledtabular}
\end{table*}

\section{\label{sec:implementation} Computational Implementation}

A practical implementation of the EMO theory requires four input parameters for each parent polymer: the HOCO and LUCO energies and the hole and electron effective masses. The accuracy of the MO energies and envelopes produced by this approach depends on the accuracy of these parameters and the method of solution of Eq. (\ref{eq:fullEM-SE}).

Ideally, the input parameters are extracted from experimental measurements. Specifically, according to Koopmans' theorem,\cite{Levine} approximations to the LUCO and HOCO can be obtained from photoelectron spectra. On the other hand, the effective masses can be obtained from the conductivities.\cite{Yu99,Davies98,Bredas90} In lieu of experimental data these parameters can also be determined computationally from the bandstructures of the parent polymers and Eq. (\ref{eq:defPEM}), which is the strategy adopted here.

The goal of this section and the next one is to demonstrate the internal consistency of the EMO approach. Hence, its predictions will be contrasted with the results of atomistic electronic-structure calculations, not with experimental data. Moreover, for this comparison to be meaningful, such calculations are performed employing the same method as the one used for determining the bandstructures. Since the co-oligomers studied are rather long, in order to avoid unnecessary complications the following simple methodology was employed.

First the geometries of the PA and PDA monomers were pre-optimized employing the semiempirical Austin Model 1 (AM1),\cite{Dewar85} as implemented in the Gaussian03 package.\cite{g03} This was accomplished by optimizing the geometry of oligomers of increasing lengths, until all the bond lengths and angles of the central monomer converged. For simplicity, all the oligomers were forced to maintain a planar configuration. In all cases, convergence was well achieved with decamers.\cite{Perdomo}

Second, taking these optimized monomers as unit cells, single-point bandstructure calculations were performed at the Extended H\"uckel (EH)\cite{Hoffmann88,Hoffmann63,Yang04} and DFT\cite{Yang04} levels. For the EH calculations the BICON-CEDiT package\cite{Brandle} was employed, adjusting the value of the $K$ parameter that appears in the bond integrals\cite{Hoffmann63} so that the calculated bandgaps agreed with the reported experimental values.\cite{Skotheim06} The optimal $K$ values turned out to be similar for both polymers. Since these parent polymers are to be used for building heterostructures, a single value of $K=2.43$ was determined to provide a good compromise.\cite{Perdomo} The resulting bandstructures are displayed in Fig.\ref{fig:2-bandsPA-PDA}. For the DFT calculations the SIESTA package\cite{Soler02} was employed, using the PBE functional and the pseudopotentials and DZP basis set implemented in it. The resulting bandstructures are not shown.

%
%

Third, the parabolic effective masses were extracted from these bandstructures, according to Eq. (\ref{eq:defPEM}). In the EH case, it should be borne in mind that the adjustment of $K$ is expected to yield good bandgap values, but is not guaranteed to produce accurate bandwidths and, consequently, effective masses. Therefore, the latter were determined more accurately by means of a real-space oligomeric extrapolation\cite{Pomogaeva08} based solely on HOMO and LUMO data for oligomers of increasing sizes,\cite{Perdomo,Mujica06} details and examples of which will be provided elsewhere.

Tables \ref{tab:table I} and \ref{tab:table II} contain the frontier CO energies and the effective masses of PA and PDA extracted from the EH and DFT bandstructures. In Fig. \ref{fig:4-band align} it can be seen that the band alignments predicted by DFT are different from the ones predicted by EH. (The EH and DFT alignments obtained are commonly referred to as 'type I' and 'type II', respectively.\cite{Yu99,Davies98,Jacak98}) The absolute HOCO and LUCO values produced by DFT should be more accurate than the ones produced by EH, whereas the opposite should be true for the bandgaps, since EH was parameterized here to reproduce the experimental bandgaps and DFT is known to underestimate these values.\cite{Yang04} Thus, at this point it cannot be reliably established which the correct alignments are. In addition, the effective masses predicted by DFT are somewhat larger than the ones predicted by EH, by at most a factor of two. In any case, what is important here is that now the performances of the EMO method employing parameters obtained with two very different levels of theory can be contrasted.

Armed with the eight EMO input parameters, the potential profiles for electrons and holes of the prototypical heterostructures were easily defined. In this connection, it is important to realize that the lengths of the constant segments in these profiles are calculated by adding the lengths in the $z$ direction of the optimized monomer structures, not by adding the lengths of the bonds. Now, the hole and electron EMO´s can be obtained by solving Eqs. (\ref{eq:fullEM-SE}) employing any of the methods available,\cite{Harrison} whose requirements of computer memory and time are practically negligible in comparison with standard atomistic electronic-structure methods. In this work, the purely numerical method described in ref. \cite{Montano} was chosen.

For the atomistic molecular electronic structures, single-point EH and DFT calculations were performed employing the implementations included in the Hyperchem v.7.01\cite{Hyperchem} and SIESTA\cite{Soler02} packages using the same AM1 pre-optimized monomer geometries, parameterization, functional and basis set as for the polymer bandstructure calculations.

\section{\label{sec:results} Results and Discussion}
\subsection{\label{subsec:EH} EH Input Parameters}

\begin{figure*}
\includegraphics[width=17.0cm]{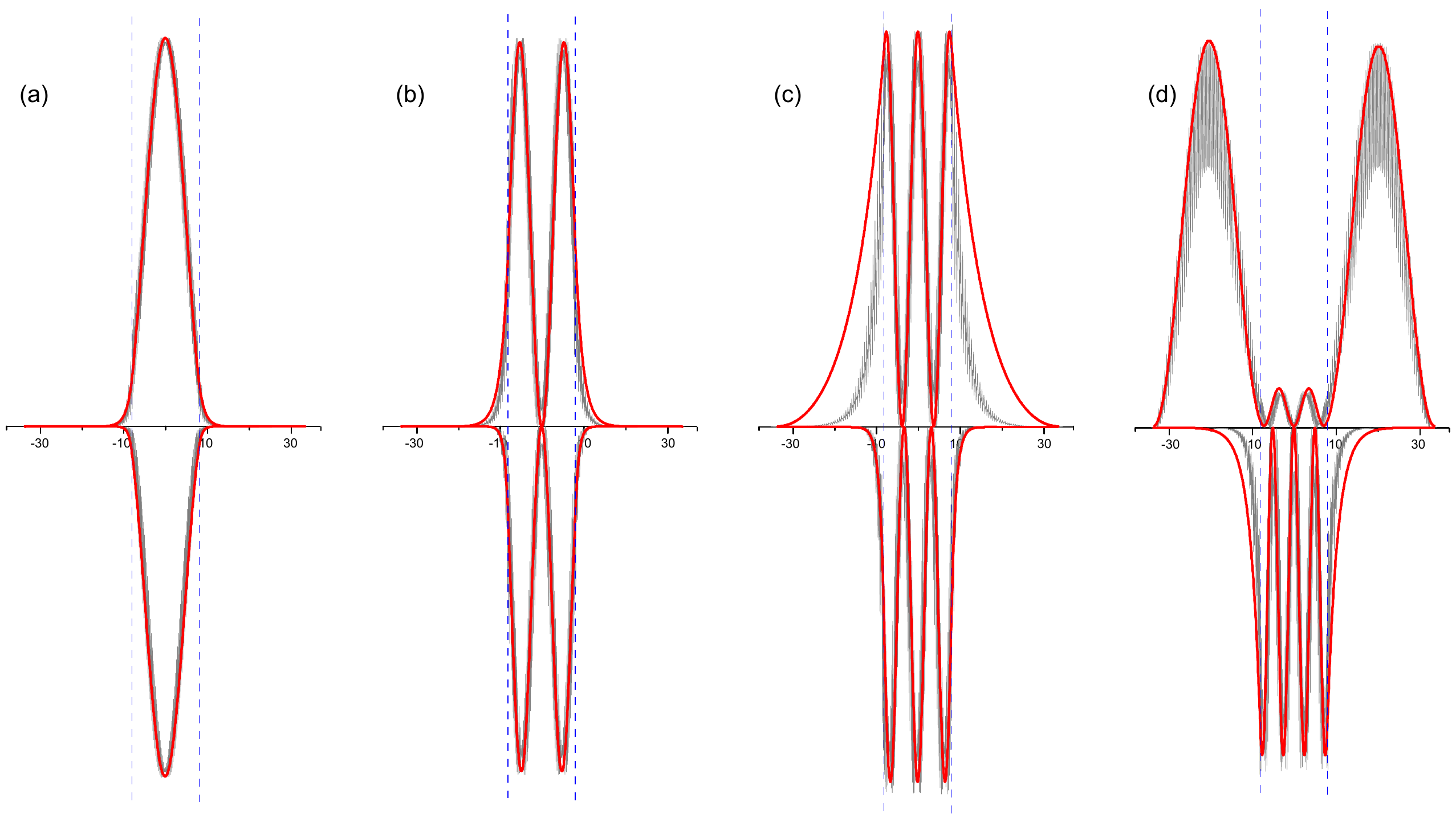}
\caption{\label{fig:5-52PDA-65PA-52PDA} Comparison of the longitudinal MO shapes (gray wiggly lines) and the EMO probability densities (red smooth lines) for (a) LUMO and HOMO, (b) LUMO+1 and HOMO-1, (c) LUMO+2 and HOMO-2, and (d) LUMO+3 and HOMO-3 of PDA$_{52}$/PA$_{65}$/PDA$_{52}$ (see
Table \ref{tab:table I}). Functions for holes are displayed ``upside down''. The vertical dashed lines indicate the positions of the intramolecular barriers. The horizontal scale is in nm.}
\end{figure*}

\begin{figure*}
\includegraphics[width=11.0cm]{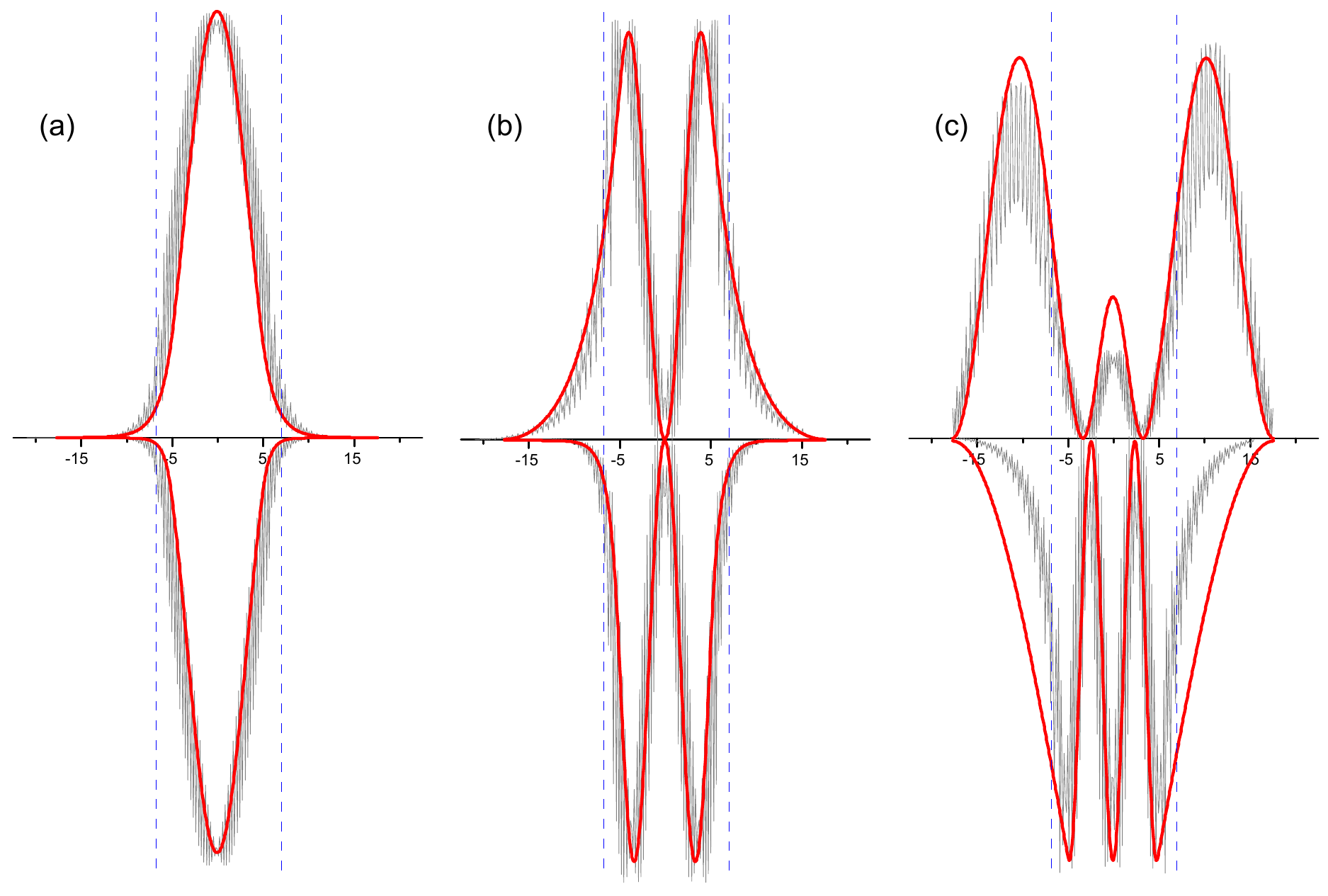}
\caption{\label{fig:6-26PDA-40PA-26PDA} Analogous to Fig. \ref{fig:5-52PDA-65PA-52PDA} for (a) LUMO and HOMO, (b) LUMO+1 and HOMO-1, and (c)
LUMO+2 and HOMO-2 of PDA$_{26}$/PA$_{40}$/PDA$_{26}$.}
\end{figure*}

In this subsection the EMO method is implemented employing the input parameters extracted from the EH calculations. The results are shown in Table \ref{tab:table I} and Figs. \ref{fig:5-52PDA-65PA-52PDA} and \ref{fig:6-26PDA-40PA-26PDA}.

First, let us focus on the prototypical three-block symmetrical structure PDA$_{52}$/PA$_{65}$/PDA$_{52}$. The potential profiles are similar to the ones shown in Fig. \ref{fig:3_scheme-PDA-PA-PDA}. The well (PA) region
and the entire molecule have lengths $L_w=16.02$ nm and $L_m=67.05$ nm, respectively. The number of bound states, $N$, for a particle of mass
$m$ confined in a symmetrical finite well of depth $V$ satisfies the inequality $N-1 < \frac{(2mV)^{1/2} L_w}{\pi \hslash} \leq N$.\cite{Levine} Employing
this formula, with the average of the hole (electron) effective masses of PA and PDA, the EMO method estimates the appearance of four (three)
states confined within the VB (CB) potential well. This is precisely what the numerical solution of Eq. (\ref{eq:fullEM-SE}) yielded. In
Table \ref{tab:table I} the energies of these states are compared with the EH MO energies. The agreement of the EMO hole (electron) energies with
the ones of the corresponding EH HOMO-3, HOMO-2, HOMO-1 and HOMO (LUMO, LUMO+1 and LUMO+2) is very good, although it increasingly deteriorates as
this variable decreases (increases). Table \ref{tab:table I} also compares the EMO energy of an electron state located slightly above the intramolecular barrier with the EH LUMO+3 energy. A very good agreement is found here too, which, interestingly, is better than for LUMO+1 and LUMO+2. The differences are of the order of $1-10$ meV, which is very encouraging.

Since the EMO's and EH MO's are one-dimensional and three-dimensional functions, respectively, they cannot be directly compared. To devise an
useful comparison criterion, the \textit{shape} of an MO along the longitudinal direction of the molecule is described in terms of the set of
parameters
\begin{equation}\label{eq:loc-par}
\rho^{(q)}_j = \sum_r | c^{(q)}_{jr} |^2,
\end{equation}
where $c^{(q)}_{jr}$ is the coefficient of the $r$-th atomic basis orbital centered at the $j$-th carbon core in the $q$-th EH MO. The contributions of the 1$s$ atomic orbitals centered at the hydrogen nuclei are not taken into account. The set of atomic indices ${j}$ and
the parameter $\rho^{(q)}_j$ can be interpreted as a discretization of the $z$ coordinate and the probability density associated with the $q$-th
MO at the $j$-th position, respectively.

Fig. \ref{fig:5-52PDA-65PA-52PDA} shows a comparison of such MO longitudinal shapes (appropriately scaled) with the EMO probability densities. It is seen, firstly, that
the EH MO's display long-wavelength large-amplitude envelopes, accompanied by short-wavelength low-amplitude oscillations that can be attributed
to the band-edge background Bloch functions, as Eqs. (\ref{eq:MOEFAv}) and (\ref{eq:MOEFAc}) predict. Secondly, the EH MO's confined within the well (PA) region exhibit
the same behavior as the square-well eigenfunctions. Thirdly, the EMO's describe very precisely the long-range variations of the EH MO's, except
for deviations of HOMO-3, LUMO+1 and LUMO+2 within the barrier regions, and LUMO+3 within the well region.

Let us assess qualitatively the contribution of each one of the main approximations involved in the theoretical development to the origins of the
observed differences in energy and spatial behavior between the EMO's and EH MO's. In the first place, the fact that no appreciable spatial deviations are found in the immediate neighborhoods of the heterojunctions indicates that the uncoupled approximation and the artificial boundary conditions work very well for this system. This is likely due to the structural similarity of PA and PDA. In the second place, the observed agreement between the EMO and EH MO wavefunctions within the classically-allowed regions and the deviations thereof within the classically-forbidden regions corroborate the line of argumentation preceding Eqs. (\ref{eq:MOEFAv}) and (\ref{eq:MOEFAc}), concerning the single-envelope
approximation. In particular, an EMO is more delocalized within the classically-forbidden regions than the corresponding EH MO because there
the latter is more accurately represented by the convolution product \cite{Arfken} of $F^{(R_{PDA})}_m (z)$ with the Fourier transform to $z$-space of $u^{(PDA)}_n (k,\vec{r})$ (see Eqs. (\ref{eq:LCCO}) and (\ref{eq:defenv})), than by the algebraic product
$F^{(R_{PDA})}_m (z) u^{(PDA)}_n (k,\vec{r})$ (Eqs. (\ref{eq:MOEFAv}) and (\ref{eq:MOEFAc})). Such convolution yields a narrower wavefunction because
$u^{(PDA)}_n (k,\vec{r})$ is confined to the first BZ in $k$-space, so its Fourier transform to $z$-space is localized within the
length of the PDA unit cell, $\ell_a$. Naturally, these deviations are more noticeable for states close to the edges of the wells than
for deeply-lying ones, since the wavefunctions of the former penetrate more into the classically-forbidden regions. In the third place, the hole and electron MO's confined inside the well have mostly PA character and their energies are seen to lie inside the parabolic
regions of the VB and CB dispersions of PA, respectively (see Table \ref{tab:table I} and Figs. \ref{fig:2-bandsPA-PDA}, \ref{fig:3_scheme-PDA-PA-PDA} and \ref{fig:4-band align}). Hence, the error introduced by the parabolic EMA is very small for these states. On the other hand, the
LUMO+3 has a dominant PDA character because it is not confined within the well. Since its energy lies slightly above the barrier, it lies
deeply within the parabolic region of the CB of PDA. The very slight deviations observed across the well region must be due to small errors in the parabolic EMA for PA, since this state is relatively far from the bottom of the well. This explains why the agreement for this state is better than for LUMO+1 and LUMO+2.

Second, let us consider the shorter three-block symmetrical structure PDA$_{26}$/PA$_{40}$/PDA$_{26}$. The well (PA) region and the entire molecule have lengths
$L_w=9.86$ nm and $L_m=35.37$ nm, respectively. Employing the same considerations as for the previous QD, the EMO method estimates the appearance
of two states confined in each potential well. Again, this is precisely what the numerical solution of Eq. (\ref{eq:fullEM-SE}) produced. Table
\ref{tab:table I} presents the energies of these states plus the energies of one additional hole state and one additional electron state located
below and above the corresponding barriers. Comparison with the energies of the corresponding EH HOMO-2, HOMO-1, HOMO, LUMO, LUMO+1 and LUMO+2,
shows qualitatively the same behavior as for the previous QD, although now the deviations tend to be about twice as large. This is expected,
since now the potential well and the molecule are shorter, causing the hole and electron states to be pushed closer to the edges of their
respective wells, and making the influence of the interfaces more noticeable. Fig \ref{fig:6-26PDA-40PA-26PDA} displays the comparison between the
EMO probability densities and the EH MO shapes. Again, the qualitative behaviors observed are analogous to the ones of the previous QD. Quantitatively, it can be seen
that the deviation of the (unconfined) LUMO+2 across the well (PA) region is larger than the one of the LUMO+3 of the previous case. This is
because the latter is closer to the LUCO of PA, so the parabolic EMA in this region works better for it. It is also interesting to notice that
the EH HOMO-2 is bound with respect to the well, while the corresponding EMO is barely unbound, lying only $2$ meV above the barrier. This effect is attributed to a small deviation from the parabolic approximation in the effective mass.

Third, let us consider the two-block asymmetrical structure PDA$_{20}$/PA$_{20}$, which, according to Fig. (\ref{fig:4-band align}), exhibits a well within the PA region. The well and the entire molecule have lengths $L_w=4.92$ mn and $L_m=14.72$ nm, respectively. Table \ref{tab:table I} shows that now only the HOMO and LUMO are confined within the well (PA) region. The degree of agreement between the EH and EMO energies for this structure is slightly lower than for PDA$_{26}$/PA$_{40}$/PDA$_{26}$, due to its shorter length, but is still very good. The EMO probability densities and the EH MO shapes (not shown) display the expected deformations associated with the asymmetry of the structure, and qualitatively agree to the same degree as in PDA$_{26}$/PA$_{40}$/PDA$_{26}$.

Finally, let us consider the one-block (single oligomer) structure PA$_{65}$, which can be visualized as the ``heterostructures"  vacuum/PA$_{65}$/vacuum, vacuum/PA$_{20}$/PA$_{25}$/PA$_{20}$/vacuum or any other appropriate partition of 65 PA monomers. A comparison of this system with the true heterostructure PDA$_{52}$/PA$_{65}$/PDA$_{52}$ helps to further assess the effects of the uncoupled approximation and the artificial boundary conditions at oligomer-oligomer and oligomer-vacuum interfaces. In Table \ref{tab:table I} it is first observed that the electron and hole levels of PA$_{65}$ have been slightly pushed up and down, respectively, in comparison with the corresponding confined levels of PDA$_{52}$/PA$_{65}$/PDA$_{52}$, which is expected because the PA-vacuum barriers are much larger than the PA-PDA barriers, causing a stronger confinement. Second, the degree of agreement between the EMO and the EH MO energies is somewhat lower than for PDA$_{52}$/PA$_{65}$/PDA$_{52}$, which makes sense because the potential and effective-mass jumps are much larger at the PA-vacuum interface than at the PA-PDA interface, so the effects of the uncoupled approximation and the artificial boundary conditions should be more noticeable.

\subsection{\label{subsec:DFT} DFT Input Parameters}

In this subsection the EMO method is implemented for the smaller structures, employing the input parameters extracted from the DFT calculations. The results are shown in Table \ref{tab:table II}. In this case, a comparison between the EMO probability densities and the Kohn-Sham orbital shapes will not be provided.

First, let us consider again the two-block asymmetrical structure PDA$_{20}$/PA$_{20}$, which, according to the DFT alignments (Fig. \ref{fig:4-band align}), now exhibits a well for electrons within the PDA region and a well for holes within the PA region. Table \ref{tab:table II} shows that according to the EMO method there are now two confined hole states within the well (PA) region, while with the EH parameters only one had been found. This comes about because the DFT offset is larger than the EH one. For electrons, three states confined within the well (PDA) region are found. The degree of agreement between the DFT and EMO energies is about the same for some states and lower for other states in comparison with the one between the EH and EMO energies of the previous subsection.

Second, let us consider the one-block (single oligomer) structure PDA$_{20}$, which can be visualized as the ``heterostructures" vacuum/PDA$_{20}$/vacuum, vacuum/PDA$_{5}$/PDA$_{10}$/PDA$_{5}$/vacuum or any other appropriate partition of 20 PDA monomers. It is seen that the level of agreement between the DFT and EMO energies is slightly lower than for PDA$_{20}$/PA$_{20}$. This is expected for reasons analogous to the ones argued for the last structure studied in the previous subsection. Furthermore, it is expected that the PDA$_{20}$ electron states should be slightly pushed up in comparison with the electron states confined within the PDA region in PDA$_{20}$/PA$_{20}$, because the PDA-vacuum barrier is much larger than the PDA-PA one, causing a stronger confinement. It is observed that the EMO results display the correct trend of behavior, whereas the DFT ones do not. This comes as no surprise, since DFT/PBE is known to be rather inaccurate for conjugated systems.\cite{Yang04} (An analogous comparison for holes is not possible because in PDA$_{20}$/PA$_{20}$, these are confined within the PA region.)

This section is closed with the following additional comments: First, the results of calculations on smaller structures (not shown) exhibited the same trends of behavior, in particular the expected deterioration with decreasing oligomer size of the agreement between the EMO and atomistic results. On the other hand, it should be evident by now that results for very large, mesoscopic heterostructures are likely to be equally, or even more, accurate, than the ones shown here. Second, no localized states at oligomer-oligomer or oligomer-vacuum interfaces were found in these structures from the atomistic calculations within the energy region examined, although it is possible that such states appear at higher or lower energies.\cite{Schwartz04,Ladik87} The uncoupled approximation and the artificial boundary conditions used in this work preclude a prediction of this kind of state.

\section{\label{sec:conclusions} Conclusions and Perspectives}

The EMO theory introduced in this paper, which constitutes an extension for (finite) molecules of the envelope-function
approximation widely employed in solid-state physics, is seen to provide an attractive conceptual and computational framework for the
description of the course-grained (non-atomistic) electronic structure of extended quasi-linear heterostructures. Its simplicity, as compared with atomistic
electronic-structure methods, stems from the mapping of the three-dimensional single-particle eigenvalue equations into one-dimensional Schr\"odinger-like equations, whose eigenfunctions describe automatically the envelopes of the MO's. The practical implementation of this
approach requires four input parameters for each parent polymer, namely the HOCO (top of valence band) and LUCO (bottom of conduction band) energies and the hole and electron effective masses (in some cases, the LOCO energies of the lateral parent polymers may also be needed),which, ideally, can be extracted from experimental measurements or, in lieu of these, from electronic bandstructure calculations. It should be born in mind that the need for these parameters arises not because the EMO approach is intrinsically a semiempirical  model, but because it is an effective theory.

As an illustration, the approach was applied to model $\pi$-conjugated co-oligomers constituting so-called quantum dots, employing input
parameters extracted from EH and DFT bandstructure calculations. To meaningfully assess the internal consistency and accuracy of the EMO approach, its predictions were compared with the results of concomitant atomistic molecular electronic-structure calculations. For high-lying holes and low-lying electrons, the agreement between the EMO and atomistic energies was very good (within the order of $1-10$ meV). In addition, the
envelope MO's were found to describe very precisely the long-wavelength oscillations of the full MO's, except when these penetrate deeply into classically-forbidden zones, where small deviations are observed. The origins of all these discrepancies were qualitatively accounted for.

The approximations involved in the formal development were introduced on heuristic grounds and validated \textit{a posteriori}. The main shortcoming resulting from these approximations is the incapability of predicting states localized at the heterojunctions. A quantitative estimation of errors and the prediction of such localized states require the examination of the couplings between bands and the reconstruction effects at the interfaces. These tasks will be undertaken in future reports.

It is of considerable interest to extend the EMO theory for the calculation of the optical\cite{Bredas90,Champagne94} and charge-transport\cite{Bredas90} properties of conjugated molecular structures. Work is under way in these directions.\cite{Serna10,Rivera11}

It would be useful also to generalize the EMO method to systems with diverse compositions and higher dimensionalities, for example graphene-based structures\cite{Rao09,Trauzettel07,Ponomarenko08} and hyper-branched, or dendritic, molecules.\cite{Lupton02,Wu06}

As far as applications are concerned, the ability to design molecular structures with properties specifically engineered for the construction of bulk or molecular electronic and optoelectronic devices is very important.\cite{Bredas90,Ellenbogen00} The EMO method may constitute a useful aid for this task, since it readily permits the tailoring of the energy levels and wavefunctions, by controlling the sizes of the constituent fragments, avoiding the costly and tedious performance of extensive electronic-structure calculations every time. For example, it has already been applied to the design of nanometric molecular architectures\cite{Mujica09} for quantum computing\cite{Trauzettel07}, molecular superlattices\cite{Mujica10} with potential applications in optoelectronics\cite{Bredas90,Lupton02,Ellenbogen00} and nanotube-based double-barrier heterostructures for chemical sensing.\cite{Rivera11}. Moreover, it is desirable to develop methodologies for the design of parent materials with bandgaps and effective masses tailored for specific applications; progress in this direction is also under way.\cite{Gil10}

There is an obvious similarity between the EMO theory, if applied to a single oligomer, and the FEMO model.\cite{Kuhn49,Platt49,Ruedenberg53} Clearly, in the former the effective box-like potentials and masses for electrons and holes arise as a matter of course from a first-principles formalism, whereas in the latter the box potential and effective mass for electrons are introduced on grounds of intuitive, albeit very clever, considerations. This lack of a rigorous derivation of the FEMO model hampers its systematic improvement. Nevertheless, now it appears that this model can, in fact, be easily deduced from the EMO theory by making a few additional approximations.

\begin{acknowledgments}
The authors are grateful to J.F. Rivera for technical assistance with the DFT calculations and to A. Restrepo for helpful discussions.
J.C.A. and C.A.M. acknowledge Colciencias for partial financial support, under contract 1106-45-221296. M.L.Z. acknowledges Direcci\'on de
Investigaciones and CICBA, Universidad Santiago de Cali, for partial financial support.
\end{acknowledgments}


\begin{thebibliography}{65}
\bibitem[{not()}]{note1}
\bibinfo{note}{Sometimes these blocks are referred to as ``quantum wells", but
  to conform to the current terminology in the semiconductor literature, it is
  more appropriate to call them ``quantum dots", since the carriers are
  confined in the three directions.}
\end{thebibliography}

\end{document}